\documentclass[runningheads]{llncs}
\usepackage{graphicx}
\usepackage{xcolor}
\usepackage{ctable}
\usepackage{multirow}
\usepackage{cleveref}
\usepackage{subcaption}
\usepackage{caption}

\graphicspath{{figures/}}
\DeclareGraphicsExtensions{.pdf,.png}

\newcommand{\etal}{\mbox{\emph{et al.}}}
\newcommand{\tabheader}[1]{%
  \begin{tabular}{@{}c@{}}
  \strut#1\strut
  \end{tabular}%
}

\begin{document}

\title{Extraction of volumetric indices from echocardiography: which deep learning solution for clinical use?}

\author{
Hang Jung Ling\inst{1}\orcidID{0000-0003-0475-9121} \and
Nathan Painchaud\inst{1,2}\orcidID{0000-0001-8269-5704} \and
Pierre-Yves Courand\inst{1,3,4}\orcidID{0000-0003-3199-7977} \and
Pierre-Marc Jodoin\inst{2}\orcidID{0000-0002-6038-5753} \and
Damien Garcia\inst{1}\orcidID{0000-0002-8552-1475} \and
Olivier Bernard\inst{1}\orcidID{0000-0003-0752-9946}
}

\authorrunning{H. J. Ling \etal}
\institute{
Univ Lyon, INSA‐Lyon, Université Claude Bernard Lyon 1, UJM-Saint Etienne, CNRS, Inserm, CREATIS UMR 5220, U1294, F‐69621, Lyon, France \and
Dept. of Computer Science, University of Sherbrooke, Sherbrooke, QC, Canada \and
Cardiology Dept., Hôpital Croix-Rousse, Hospices Civils de Lyon, Lyon, France \and
Cardiology Dept., Hôpital Lyon Sud, Hospices Civils de Lyon, Lyon, France
\vspace{-0.5cm}}
\maketitle              %
\begin{abstract}
Deep learning-based methods have spearheaded the automatic analysis of echocardiographic images, taking advantage of the publication of multiple open access datasets annotated by experts (CAMUS being one of the largest public databases). %
However, these models are still considered unreliable by clinicians due to unresolved issues concerning i) the temporal consistency of their predictions, and ii) their ability to generalize across datasets. In this context, we propose a comprehensive comparison between the current best performing methods in medical/echocardiographic image segmentation, with a particular focus on temporal consistency and cross-dataset aspects. We introduce a new private dataset, named CARDINAL, of apical two-chamber and apical four-chamber sequences, with reference segmentation over the full cardiac cycle. We show that the proposed 3D \mbox{nnU-Net} outperforms alternative 2D and recurrent segmentation methods. We also report that the best models trained on CARDINAL, when tested on CAMUS without any fine-tuning, still manage to perform competitively with respect to prior methods. Overall, the experimental results suggest that with sufficient training data, 3D \mbox{nnU-Net} could become the first automated tool to finally meet the standards of an everyday clinical device.

\end{abstract}
\vspace{-0.5cm}
\section{Introduction}

Echocardiographic imaging has undergone major advances in recent years thanks to artificial intelligence, especially the deep learning (DL) paradigm. In particular, the automated extraction of clinical indices from the segmentation of cardiac structures has been the subject of intense research leading to major breakthroughs. A key component of these advances has been the publication of open access annotated datasets, including CETUS (45 patients, 3D images annotated at End-Diastole - ED and End-Systole - ES) \cite{bernard_challenge_2014}, CAMUS (500 patients, 2D images annotated at ED and ES in apical two-chamber - A2C - and apical four-chamber - A4C - views) \cite{leclerc_deep_2019}, EchoNet-Dynamic (10,036 patients, 2D sub-sampled images annotated at ED and ES in A4C view) \cite{ouyang_video-based_2020}, HMC-QU (109 patients, 2D sequences annotated in A4C view) \cite{degerli_early_2021} and TED (98 patients from the CAMUS dataset, 2D sequences annotated in A4C view) \cite{painchaud_echocardiography_2022}.

These datasets allowed effective and fair comparisons of methods, whether they are generic image segmentation models \cite{leclerc_deep_2019,ling_ius_2022} or were specifically designed to process echocardiographic images \cite{wei_temporal-consistent_2020,sfakianakis_gudu_2023}. Thus, the performance of current state-of-the-art (SOTA) methods on the CAMUS dataset confirmed the dominance of the DL-based methods, which finally achieved inter- and intra-observer variability for most of the geometric (Dice score, Hausdorff distance, mean absolute distance - MAD) and clinical metrics (ejection fraction - EF, volumes at ED/ES).

Although these results are extremely promising and represent a crucial step towards the automation of echocardiographic image analysis, they are not sufficient to justify confidence in fully automated methods in a clinical context. Indeed, two crucial challenges on the path to the practical application of these algorithms remain understudied in the field: i) the frame-by-frame temporal consistency of the predictions, and ii) the generalization of the methods across datasets. Based on this observation, we propose the following contributions:

\begin{enumerate}
    \item  We study the performance of two generic architectures based on common temporal data processing techniques on 2D echocardiography sequences, and compare them to current SOTAs in the same field;
    \item We present a new private dataset called CARDINAL (240 patients, 2D sequences annotated in A4C and A2C views), and report the performance impact of training our methods exclusively on CARDINAL and testing on CAMUS.
\end{enumerate}

\vspace{-0.2cm}
\section{Benchmarked Methods}
\label{sec:benchmarked-methods}

CAMUS is currently the only dataset where an evaluation platform has been established to effectively compare the performance of segmentation methods\footnote{\url{https://www.creatis.insa-lyon.fr/Challenge/camus/results.html}}. We therefore relied on this dataset to select the methods we retained in this study.

\vspace{-0.2cm}
\subsection{2D DL Methods}
The currently best performing method on the CAMUS dataset exploits the \mbox{nnU-Net} formalism~\cite{ling_ius_2022}. This model is based on the \mbox{U-Net} architecture and implements several successful DL tricks, such as a patch-wise approach to preserve image resolution, data augmentation during both training and inference to enforce generalization and automatic hyperparameter search of the \mbox{U-Net} architecture to increase accuracy. Note that the 2D version uses only one \mbox{U-Net} model~\cite{isensee_nnu-net_2021}.

Recently, Sfakianakis \etal~\cite{sfakianakis_gudu_2023} developed a DL solution called GUDU based on three key aspects. First, they proposed to use data augmentations tailored to ultrasound acquisition, i.e. variation of the contrast between the myocardial tissue and the left ventricular (LV) cavity, random rotation from the origin of the sectorial shape to mimic different probe positioning, and perspective transformations to simulate probe twisting. Inspired by ensemble models, the authors also trained 5 U-Nets with different architectures and averaged their outputs during inference in order to compute the final prediction. Finally, a new loss function was proposed that takes into account the relative position of the cardiac structures with respect to each other. The authors demonstrated the usefulness and complementarity of each contribution in an ablation study.

\subsection{2D+t DL Methods}
Despite the success of 2D methods in producing accurate segmentations for individual echocardiographic frames, they often fail to maintain temporal consistency between frames~\cite{painchaud_echocardiography_2022}. Prior to the publication of HMC-QU and TED, there were no publicly available datasets to train and compare methods that incorporate the temporal dimension in 2D+time echocardiography. This explains why so few papers have focused on this topic.

Due to the lack of 2D+time annotated datasets, Wei \etal~\cite{wei_temporal-consistent_2020} proposed a method to leverage the limited ED/ES annotated frames and propagate them to unannotated frames. This was achieved by training a 3D \mbox{U-Net} designed to predict both the deformation fields between each pair of consecutive frames and the segmentation masks at each frame of the sequence. The deformation fields are used to propagate the ED/ES reference annotations forward and backward in time through the sequence. The corresponding propagated masks are then used as targets for self-supervised segmentation of the entire sequence. This encourages the model to learn consistent temporal dynamics to find the best match between the predicted segmentation masks and the propagated annotations.

More recently, Smistad \etal~\cite{smistad_ius_2021} and Hu \etal~\cite{hu_ius_2022} added convolutional long-term memory blocks to each layer of the encoder of a 2D \mbox{U-Net} (this type of model is hereafter referred to as \mbox{U-Net} LSTM). Thus, instead of processing a single frame, these methods take a series of frames as input and store the extracted features over time to produce the final segmentation of the entire sequence. Results show that such a strategy tends to reduce segmentation shifts from one frame to another.

By their very nature, echocardiographic sequences exhibit regular properties along the time axis. Therefore, it seems logical to consider 2D ultrasound sequences as complete volumes containing coherent 3D shapes and to extract 3D features using 3D convolutional layers to promote temporal consistency. Thus, in this paper, we propose to train a 3D \mbox{nnU-Net} to segment the complete cardiac sequences in a single run. We hypothesize that this model will inherently learn temporal consistency while maintaining a high level of segmentation accuracy.

\section{Experimental Setup}

\subsection{CARDINAL Dataset}

\subsubsection{Acquisition Protocol:}
The proposed dataset consists of clinical examinations of 240 patients, acquired at the University Hospital of Lyon (Croix-Rousse Lyon Sud, France) under the regulation of the local ethics committee of the hospital. The complete dataset was acquired with GE ultrasound scanners. %
For each patient, 2D A4C and A2C view sequences were exported from the EchoPAC analysis software. %
Each exported sequence corresponds to a set of B-mode images expressed in polar coordinates. The same interpolation procedure as used for the CAMUS dataset was applied to express all sequences in Cartesian coordinates with a single grid resolution of 0.31 mm$^2$. %
Each sequence in the CARDINAL dataset corresponds to a complete cardiac cycle defined as the interval between peaks of maximal LV cavity surface area.

\subsubsection{Reference Annotations:}
To tackle the total number of frames to be annotated, an experienced observer first delineated the different contours using semi-automatic tools to ensure temporal consistency of the segmented shapes. Each corresponding output was then checked/corrected by two other experienced observers. To identify the ED/ES frames in the sequence, the ED frames correspond (by definition) to the beginning and end of each sequence, and the ES frame corresponds to the frame where the LV cavity surface is smallest.

\subsection{Implemented DL Methods}
For a fair comparison, we implemented the 2D \mbox{nnU-Net}, \mbox{U-Net} LSTM, and 3D \mbox{nnU-Net} described in \cref{sec:benchmarked-methods} using the same Python library called ASCENT \footnote{\url{https://github.com/creatis-myriad/ASCENT}}. These models shared the following training hyperparameters: batch size of 2, SGD optimizer with a learning rate of 0.01 coupled with a polynomial decay scheduler, and 1000 training epochs. The 2D and 3D \mbox{nnU-Net} used a patch-wise approach to avoid resizing the images, thus preserving the native image resolution. To train the \mbox{U-Net} LSTM, the input images were resized to \mbox{$256 \times 256$} and 24 consecutive frames were randomly selected and fed to the model to produce the corresponding segmentations. For inference, the sliding window approach with a Gaussian importance map was used. The prediction was given by the average of the \textit{softmax} probabilities of all windows. To improve segmentation accuracy, the final prediction was obtained by averaging the predictions of the original and mirrored images along different axes. More implementation details for each model can be found in \cref{tab:models_details}.

\begin{table*}[ht]
    \centering
    \caption{Details of the implementation of the three methods evaluated in this study. \textit{Lowest resolution}: Size of the lowest resolution of feature maps in pixels. \textit{Optimization scheme}: Optimizer + initial learning rate + learning rate scheduler used. \textit{Training duration (hours)}: number of hours required to train each model for 1000 epochs. The configurations shared between models are only shown once in their respective rows.}
    \begin{tabular*}{\textwidth}
       {@{} @{\extracolsep{\fill}} l c c c @{}}
        \toprule
        Configurations & 3D nnU-Net & 2D nnU-Net & U-Net LSTM\\
        \midrule
        Patch size (pixels) & $320 \times 256 \times 24$ & $640 \times 512$ & $256 \times 256 \times 24$\\
        Batch size & \multicolumn{3}{c}{2}\\
        \midrule
        Nb. feature maps & \multicolumn{3}{c}{$32\downarrow480\uparrow32$}\\
        Lowest resolution & $10 \times 8 \times 6$ & $5 \times 4$ & $8 \times 8 \times 24$\\
        \midrule
        Downsampling scheme & \multicolumn{3}{c}{Stride pooling}\\
        Upsampling scheme & \multicolumn{3}{c}{Deconvolution}\\
        Normalization scheme & \multicolumn{3}{c}{Instance normalization}\\
        \midrule
        Optimization scheme & \multicolumn{3}{c}{SGD + 0.01 + polynomial decay}\\
        Loss function & \multicolumn{3}{c}{Cross entropy + Dice}\\
        \midrule
        Number of parameters & 41.3 M & 30.4 M & 49 M\\
        Training duration (hours) & 22.8 & 8 & 69.5\\
        \bottomrule \vspace{-0.8cm}
    \end{tabular*}
\label{tab:models_details}
\end{table*}

\section{Results}
We evaluate the methods described in \cref{sec:benchmarked-methods} using three types of measures to get a complete picture of their performance in terms of segmentation accuracy (\cref{tab:segmentation_metrics}), extraction of clinical indices (\cref{tab:clinical_metrics}) and temporal consistency (\cref{tab:temporal_metrics}). In each of these tables, we group the methods according to the datasets on which they were trained and tested (CARDINAL is abbreviated as \textit{CL} and CAMUS is abbreviated as \textit{CS}) to make it easier to observe the change in performance when generalizing to a new dataset.

\subsection{Geometric and Clinical Accuracy}
\Cref{tab:segmentation_metrics} shows the segmentation accuracy computed from the CARDINAL and CAMUS datasets for the 5 algorithms described in \cref{sec:benchmarked-methods}. The values in bold correspond to the best scores for each metric for a given training/test dataset setup. From the results on the CARDINAL dataset (CL/CL case), we can see that the 3D \mbox{nnU-Net} has the best segmentation scores for all metrics, for both ED and ES. It is also interesting to note that the two temporal consistency methods (3D \mbox{nnU-Net} and \mbox{U-Net} LSTM) produce better results than the 2D \mbox{nnU-Net} method. This can be explained by the fact that the reference segmentation has regular properties along the temporal axis due to the annotation process. Methods that integrate the temporal dimension into their architecture are therefore more likely to produce segmentation results that are closer to the manual references. 

It is worth mentioning that methods trained and tested on the same dataset (sections CL/CL and CS/CS in \Cref{tab:segmentation_metrics}) get overall better results up to $1.7x$ for the Hausdorff and MAD metrics.  One reason for such an improvement is the larger amount of annotated training images for CARDINAL (18,793 images from 190 training/validation patients, reference frames for the full cardiac cycle in A2C and A4C views) than for CAMUS (1,800 images from 450 training/validation patients, reference frames at ED and ES in A2C and A4C views). 

\Cref{tab:clinical_metrics} reports the clinical metrics for the 5 methods. As in \Cref{tab:segmentation_metrics}, the methods enforcing temporal consistency gets the best results on CARDINAL, especially for the ejection fraction for which temporal consistency is essential (mean correlation score of $0.917$).  Furthermore, the best models trained on CARDINAL or CAMUS produce similar results for volume estimation (average correlation of $0.978$), revealing a limit reached by these approaches, certainly due to the resolution of the imaging systems.

\begin{table*}[ht]
    \centering
    \caption{LV segmentation accuracy of the benchmarked methods, on different subsets of frames. The columns \textit{All}, \textit{ED} and \textit{ES} indicate results averaged over all frames, only ED frames, and only ES frames, respectively. Since CAMUS only provides annotation for ED/ES frames, results over \textit{all} frames are not available when testing on it. \\{\scriptsize (CL:CARDINAL, CS:CAMUS)}}
    \begin{tabular*}{\textwidth}
        {@{} @{\extracolsep{\fill}} l l ccc ccc ccc @{}}
        \toprule
        \multirow{3}*{Methods} & \multirow{3}*{Train/test} & \multicolumn{3}{c}{Dice} & \multicolumn{3}{c}{Hausdorff (mm)} & \multicolumn{3}{c}{MAD (mm)} \\
        \cmidrule(lr){3-5} \cmidrule(lr){6-8} \cmidrule(lr){9-11} & & \multicolumn{1}{c}{All} & \multicolumn{1}{c}{ED} & \multicolumn{1}{c}{ES} & \multicolumn{1}{c}{All} & \multicolumn{1}{c}{ED} & \multicolumn{1}{c}{ES} & \multicolumn{1}{c}{All} & \multicolumn{1}{c}{ED} & \multicolumn{1}{c}{ES} \\
        \midrule
        3D nnU-Net & & \textbf{.969} & \textbf{.968} & \textbf{.960} & \textbf{2.3} & \textbf{2.7} & \textbf{2.5} & \textbf{0.7} & \textbf{0.8} & \textbf{0.7} \\
        2D nnU-Net & CL/CL & .957 & .961 & .942 & 2.9 & 3.1 & 3.1 & 0.9 & 1.0 & 1.1 \\
        U-Net LSTM & & .964 & .964 & .956 & 2.5 & 2.8 & 2.6 & 0.8 & 0.9 & 0.8 \\
        \midrule
        3D nnU-Net & & - & \textbf{.939} & \textbf{.926} & - & 5.2 & \textbf{4.6} & - & \textbf{1.6} & \textbf{1.5} \\
        2D nnU-Net & CL/CS & - & .934 & .921 & - & \textbf{4.9} & 4.6 & - & 1.8 & 1.6 \\
        U-Net LSTM & & - & .925 & .903 & - & 6.0 & 5.8 & - & 2.1 & 2.1 \\
        \midrule
        2D nnU-Net & & - & \textbf{.952} & \textbf{.935} & - & \textbf{4.3} & \textbf{4.2} & - & \textbf{1.3} & \textbf{1.3} \\
        CLAS & CS/CS & - & .947 & .929 & - & 4.6 & 4.6 & - & 1.4 & 1.4  \\
        GUDU & & - & .946 & .929 & - & 4.7 & 4.7 & - & 1.4 & 1.4 \\
        \bottomrule
    \end{tabular*}
\label{tab:segmentation_metrics}
\end{table*}

\begin{table*}[ht]
    \centering
    \caption{Clinical metrics of the benchmarked methods. The ED/ES volumes were computed from both the predicted and reference masks using Simpson's biplane method. \textit{Corr.}: Correlation between the ejection fraction (EF) derived from the predicted/reference segmentation. \textit{MAE}: Mean Absolute Error between the EF derived from predicted/reference segmentation.}
    \begin{tabular*}{\textwidth}
        {@{} @{\extracolsep{\fill}} l l cc cc cc @{}}
        \toprule
        \multirow{3}*{Methods} & \multirow{3}*{Train/test} & \multicolumn{2}{c}{EF} & \multicolumn{2}{c}{Volume ED} & \multicolumn{2}{c}{Volume ES} \\
        \cmidrule(lr){3-4} \cmidrule(lr){5-6} \cmidrule(lr){7-8} & & \multicolumn{1}{c}{Corr.} & \multicolumn{1}{c}{MAE (\%)} & \multicolumn{1}{c}{Corr.} & \multicolumn{1}{c}{MAE (ml)} & \multicolumn{1}{c}{Corr.} & \multicolumn{1}{c}{MAE (ml)} \\
        \midrule
        3D nnU-Net & & .913 & 2.9 & \textbf{.978} & \textbf{3.3} & \textbf{.974} & \textbf{2.7} \\
        2D nnU-Net & CL/CL  & .850 & 3.8 & .967 & 4.4 & .957 & 3.2 \\
        U-Net LSTM & & \textbf{.922} & \textbf{2.7} & .973 & 3.4 & .969 & 2.8 \\
        \midrule
        3D nnU-Net & & \textbf{.869} & \textbf{5.3} & \textbf{.974} & \textbf{9.6} & \textbf{.976} & \textbf{4.9} \\
        2D nnU-Net & CL/CS & .810 & 7.0 & .970 & 12.8 & .959 & 6.2 \\
        U-Net LSTM & & .822 & 11.1 & .879 & 15.9 & .903 & 8.2  \\
        \midrule
        2D nnU-Net & & .857 & 4.7 & \textbf{.977} & \textbf{5.9} & \textbf{.987} & \textbf{4.0} \\
        CLAS & CS/CS & \textbf{.926} & \textbf{4.0} & .958 & 7.7 & .979 & 4.4  \\
        GUDU & & .897 & \textbf{4.0} & \textbf{.977} & 6.7 & .981 & 4.6 \\
        \bottomrule
    \end{tabular*}
\label{tab:clinical_metrics}
\end{table*}

\subsection{Integration of Temporal Consistency}
\begin{figure}[h!]
    \centering{
    \begin{subfigure}[b]{.62\columnwidth}
        \includegraphics[width=\textwidth]{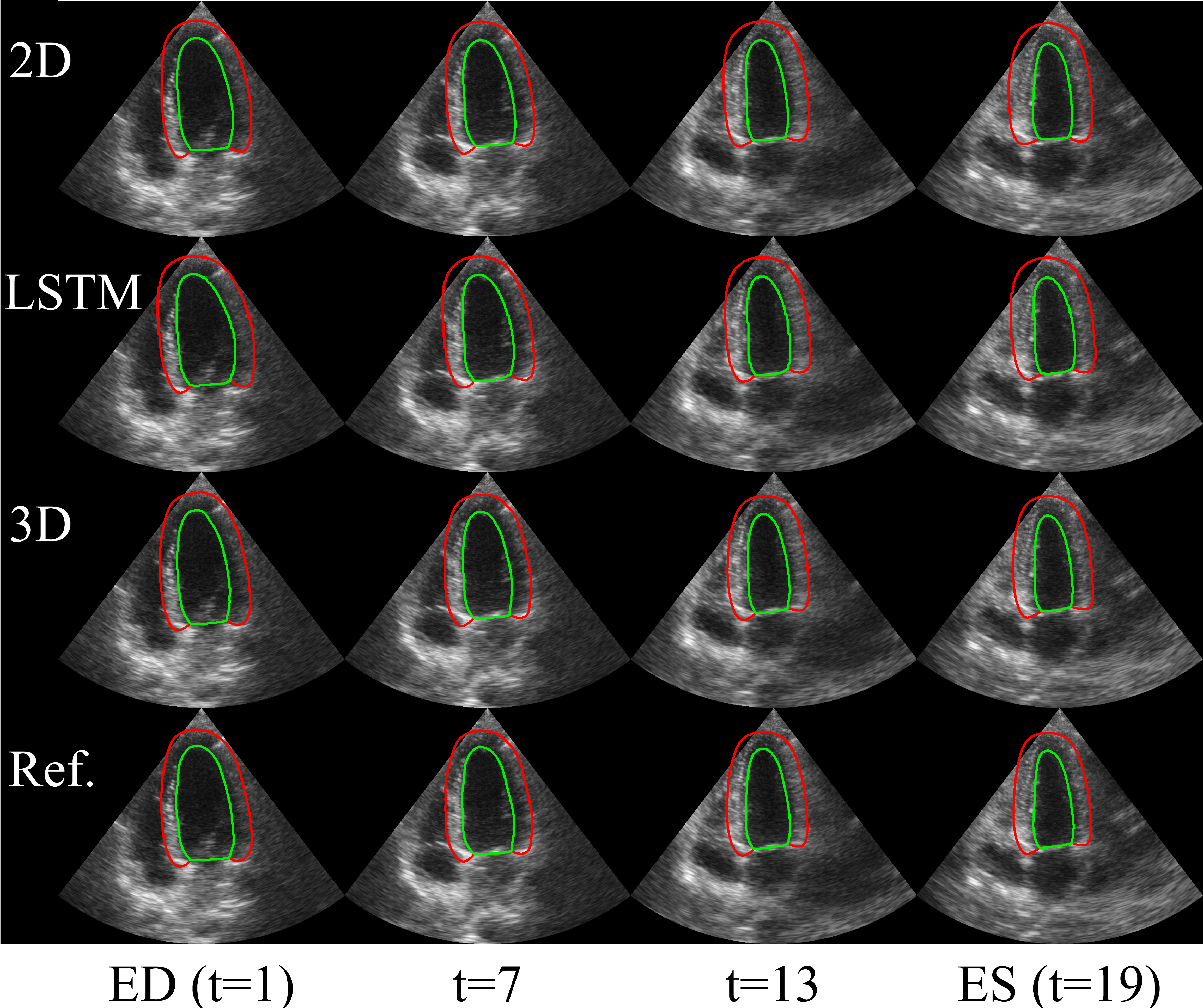}
    \end{subfigure}
    }
    \hfill
    \begin{subfigure}[b]{.36\columnwidth}
        \includegraphics[width=\textwidth]{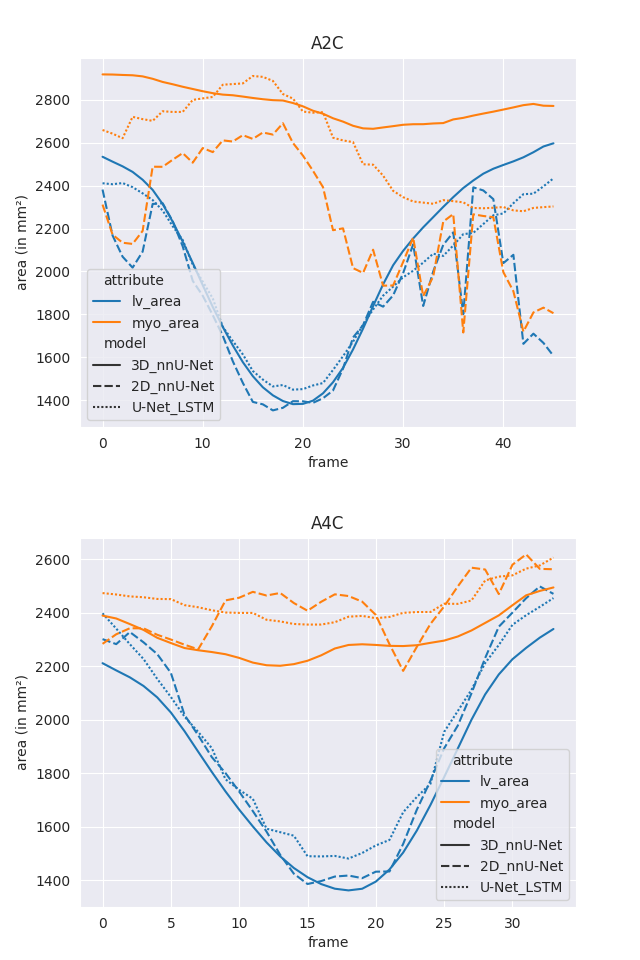}
    \end{subfigure}
    \caption{Visualization of the temporal consistency of the segmentations on one patient from the CARDINAL test set. (left) Frames sampled between ED and ES, with segmentation masks from our own methods + reference. (right) Curves of the LV and myocardium surfaces w.r.t. frame in the sequence.}
    \label{fig:temporal_figures}
\end{figure}

\Cref{tab:temporal_metrics} allows a better investigation of the temporal performance of the methods by providing additional information on the number/percentage of frames  considered temporally inconsistent w.r.t. their neighboring frames. As expected, the methods incorporating temporal persistence produced fewer temporal errors. Looking at the number of sequences with at least one temporally inconsistent frame, the 3D \mbox{nnU-Net} clearly outperforms \mbox{U-Net} LSTM, with only $4$ inconsistent sequences over $100$ compared to $98$ sequences for \mbox{U-Net} LSTM. This result illustrates the greater ability of features computed from 3D convolutional layers to extract relevant spatio-temporal information. The few remaining temporal errors for the 3D \mbox{nnU-Net} are more an indication that the metrics we used are (overly) strict on the temporal smoothness. %
Indeed, the 3D \mbox{nnU-Net} temporal "inconsistencies" appear invisible to the expert eye. As a qualitative evaluation, \cref{fig:temporal_figures} illustrates in detail the temporal consistency of each of our own method on one patient from the CARDINAL test set. To complement this, we also provide in the supplementary material examples of temporally consistent and inconsistent segmentation results obtained by the 3D \mbox{nnU-Net} method for the CARDINAL and CAMUS datasets.

\begin{table*}[ht]
    \centering
    \caption{Temporal consistency of the benchmarked methods, as defined in~\cite{painchaud_echocardiography_2022}.
    \textit{Nb of seq. w/ err.}: number of sequences (out of the 100 testing sequences) where at least one frame is temporally inconsistent. \textit{\% of frames w/ err.}: percentage of frames that are inconsistent in the sequences with at least one temporally inconsistent frame. \textit{Err. to thresh. ratio}: average ratio between the measure used to identify temporal inconsistencies and the threshold for temporal inconsistencies. A lower value indicates ``smoother'' temporal segmentations.} %
    \begin{tabular*}{0.9\textwidth}
        {@{} @{\extracolsep{\fill}} l l c c c @{}}
        \toprule
        Methods & Train/test & \tabheader{Nb of seq. \\w/ err.} &\tabheader{\% of frames \\w/ err.} & \tabheader{Err. to thresh. \\ratio} \\
        \midrule
        3D nnU-Net & & \textbf{4} & \textbf{4} & \textbf{.045} \\
        2D nnU-Net & CL/CL & 100 & 30 & .210 \\
        U-Net LSTM & & 98 & 13 &  .110 \\
        \midrule
        3D nnU-Net & & \textbf{28} & \textbf{12} & \textbf{.095} \\
        2D nnU-Net & CL/CS & 85 & 21 & .162 \\
        U-Net LSTM & & 83 & 16 & .114 \\
        \bottomrule
    \end{tabular*}
\label{tab:temporal_metrics}
\end{table*}

\subsection{Generalization Across Datasets}
The ability to generalize across datasets is crucially important to gauge the capacity of a method to properly analyze data affected by a distributional shift. To this end, the models trained on CARDINAL were also evaluated on the CAMUS test set without any fine-tuning. The results are reported in the ``CL/CS'' sections of \cref{tab:segmentation_metrics,tab:clinical_metrics,tab:temporal_metrics}. Among the methods evaluated, 3D \mbox{nnU-Net} is the undisputed best.  It even  produces competitive geometric and clinical scores compared with SOTA methods trained directly on CAMUS. Thanks to the integration of temporal consistency, the 3D \mbox{nnU-Net} trained on CARDINAL also produces one of the best correlation scores for the EF calculated on the CAMUS dataset, even when compared to SOTA methods trained directly on CAMUS. In view of these results, and considering that the annotation process between the two databases was not identical and was carried out by different experts (which inevitably introduces a bias during the learning phase), the generalization capacity of the 3D \mbox{nnU-Net} model seems remarkable.

\section{Conclusion}
We evaluated the ability of different methods to accurately segment echocardiographic images, with a focus on temporal consistency and cross-dataset generalization. To this end, we introduced a new private database called CARDINAL, with annotations from an expert on the full cardiac cycle for each sequence. The results show that 3D \mbox{nnU-Net} and \mbox{U-Net} LSTM produce the best geometric and clinical scores on the CARDINAL dataset due to the integration of temporal persistence. Regarding the temporal consistency metrics, 3D \mbox{nnU-Net} performed significantly better than \mbox{U-Net} LSTM with only four sequences (instead of 98) out of 100 having at least one image that was temporally inconsistent. As far as cross-dataset generalization is concerned, 3D \mbox{nnU-Net} is also the best performing method. When trained on CARDINAL and tested on CAMUS, it achieved comparable geometric and clinical scores to the best methods both trained and tested on CAMUS. All these results clearly show that 3D \mbox{nnU-Net} is a serious candidate to become the first automated tool to meet the requirements of routine clinical examinations.

\section*{Acknowledgment}
This research was funded, in whole or in part, by l’Agence Nationale de la Recherche (ANR), project ANR-22-CE45-0029-01. The authors gratefully acknowledge financial support of the MEGA doctoral school (ED 162), the NSERC Canada Graduate Scholarships-Doctoral Program, the FRQNT Doctoral Scholarships Program, and the LABEX PRIMES (ANR-11-LABX-0063) of Université de Lyon, within the program “Investissements d’Avenir” operated by the French ANR. The authors also thank GENCI-IDRIS for providing access to HPC resources (Grant 2022-[AD010313603]). For the purpose of open access, the authors have applied a CC BY public copyright license to any Author Accepted Manuscript (AAM) version arising from this submission.

\bibliographystyle{splncs04}
\bibliography{camera-ready-FIMH-2023}

\end{document}